\documentclass{czjphyse}
\usepackage{epsf}  
\begin{document}
\title{SPIN EFFECTS IN DIFFRACTIVE
  $Q \bar Q$ PRODUCTION AND COMPASS EXPERIMENT}        
\authori{S.V.Goloskokov}      
\addressi{Bogoliubov Laboratory of Theoretical  Physics,
 Joint Institute for Nuclear Research,
Dubna 141980, Moscow region, Russia}     
\authorii{}     
\addressii{}    
\headtitle{ SPIN EFFECTS IN DIFFRACTIVE $Q \bar Q$ PRODUCTION
   \ldots}            
\headauthor{ S.V.Goloskokov}           
\specialhead{ S.V.Goloskokov: SPIN EFFECTS IN DIFFRACTIVE $Q \bar Q$
PRODUCTION\ldots} 
\evidence{A}
\daterec{XXX}    
\cislo{0}  \year{1999}
\setcounter{page}{1}
\pagesfromto{000--000}
\maketitle

\begin{abstract}
We calculate the double spin asymmetry of the diffractive $Q
\bar Q$ and vector meson leptoproduction at COMPASS energies. We
analyze dependences of the asymmetry on the structure of the
Pomeron-proton coupling. It is shown that it is difficult to
study the spin structure of the Pomeron coupling with the proton
from the $A_{ll}$ asymmetry. The $A_{lT}$ asymmetry might be an
appropriate object for this investigation.
\end{abstract}

\section{Introduction}
Now the Pomeron nature is a problem of topical interest
due to the progress in analysis of diffractive processes at HERA
\cite{h1_zeus,diff}. Study of the diffractive  vector meson
and  $Q \bar Q$ production
\begin{equation}
\label{de}
e+p \to e'+p'+Q \bar Q
\end{equation}
should give information on the gluon distribution at
small $x$ \cite{rysk,j-psi} and  on the Pomeron structure
\cite{roy}. The experiments with polarized beams should be used
to extract the information on polarized parton
distributions. The diffractive $J/\Psi$ and heavy $Q \bar Q$
production play a significant role here. The $Q \bar Q$ exchange
in $t$-channel is not essential in such processes and the
predominant contribution is determined by a color singlet $t$
-channel exchange (Pomeron).

The $A_{ll}$ asymmetry of the open charm production has been
proposed to be used by COMPASS \cite{compass}  to study $\Delta
G$. To determine a possible role of diffractive events at
COMPASS energies, the $A_{ll}$ asymmetry of the polarized
diffractive $c \bar c$ production  has been studied in
\cite{gola_ll}.  The model where the Pomeron interacts with a
single quark in the loop has been used. The diffractive
asymmetry has been found not small in this model. The ratio of
diffractive and total events at small $Q^2$ is about 30\%
\cite{diff}. The resulting diffractive contribution to $A_{ll}$
asymmetry  might be about 5-10\%. The analysis of the final jet
kinematics shows that  the $c \bar c$ production through the
photon-gluon fusion and diffractive events should be
detected simultaneously  by COMPASS. It was mentioned on the
basis of these model results \cite{gola_ll} that the diffractive
contribution might be an important background in the COMPASS
experiment. In the present report, the polarized cross section
of diffractive hadron leptoproduction at high energies will
be studied within a QCD two-gluon model of the Pomeron. We shall
investigate the polarized diffractive $Q \bar Q$ production at
COMPASS. The spin effects in diffractive vector meson production
will also be discussed shortly.

\section{Structure of hadron leptoproduction in QCD model}
Let us study the diffractive hadron leptoproduction. For the
open charm production, the leading $t$-channel contribution is
determined by the  Pomeron exchange. In the Donnachie-Landshoff
(D-L) model \cite{lansh-m}, the Pomeron preferably couples to a
single quark in the hadron (Fig. 1a). In the QCD-inspired
models, the Pomeron is presented  by two gluons \cite{low} which
can couple to a different quark in the loop as well as to  the
single one (Fig 1b, c).
\begin{figure}
\centering
\mbox{\epsfxsize=90mm\epsffile{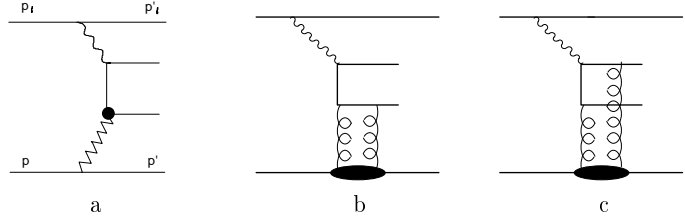}}
\caption{Donnachie-Landshoff and two-gluon model of the Pomeron}
\label{F1}
\end{figure}

The  diffractive lepton-proton reactions (\ref{de}) are usually described
 in terms of the kinematic variables
\begin{eqnarray}
\label{kvar}
Q^2=-q^2,\;t=r^2, \phantom{aaaaaaaaaa} \nonumber
\\  y=\frac{pq}{p_l p},\;x=\frac{Q^2}{2pq},\;
x_p=\frac{q(p-p')}{qp},\;\beta=\frac{x}{x_p},
\end{eqnarray}
where $p_l, p'_l$ and $p, p'$ are the initial and final lepton and
proton momenta, respectively, $q=p_l-p'_l,$ $r=p-p'$ are the virtual
photon and Pomeron momenta.

 The cross section of the hadron
leptoproduction can be decomposed into the leptonic and hadronic
tensors, the amplitude of the $\gamma^\star I\hspace{-1.6mm}P
\to Hadrons$ transition amplitude and the Pomeron exchange.
 The
structure of the leptonic tensor is quite simple \cite{efrem}
because the lepton is a point-like object
\begin{eqnarray}
\label{lept}
{\cal L}^{\mu \nu}(s_l)&=& \sum_{spin \ s_f} \bar u(p_l',s_f)
\gamma^{\mu} u(p_l,s_l) \bar u(p_l,s_l) \gamma^{\nu}
u(p_l',s_f)
\nonumber \\
&=& {\rm Tr} \left [  (/\hspace{-1.7mm} p_l+\mu) \frac{1+
\gamma_5 /\hspace{-2.1mm} s_l}{2}
\gamma^{\nu} ( /\hspace{-1.7mm} p_l'+\mu) \gamma^{\mu}
 \right] .
\end{eqnarray}
Here $s_l$ is a spin vector of the initial lepton. The
spin-average and spin dependent cross sections with parallel and
antiparallel longitudinal polarization of a lepton and a proton
are determined by the relation
\begin{equation}
\label{spm}
\sigma(\pm) =\frac{1}{2} \left( \sigma(^{\rightarrow} _{\Leftarrow}) \pm
\sigma(^{\rightarrow} _{\Rightarrow})\right).
\end{equation}
These cross sections can be expressed in terms of the
spin-average and spin dependent value of the lepton and hadron
tensors. For the first, one can write
\begin{equation}
\label{l+-}
{\cal L}^{\mu \nu}(\pm)=\frac{1}{2}({\cal L}^{\mu \nu}(+\frac{1}{2})
\pm {\cal L}^{\mu \nu}(-\frac{1}{2})).
\end{equation}
For longitudinal polarization, ${\cal L}^{\mu;\nu}(\pm\frac{1}{2})$
are the tensors with the helicity of the initial  lepton equal to
$\pm 1/2$. The tensors (\ref{l+-}) look like
\begin{eqnarray}
\label{lpm}
{\cal L}^{\mu \nu}(+)&=& 2 (g^{\mu \nu} l \cdot q + 2 l^\mu l^\nu -
l^\mu q^\nu - l^\nu q^\mu),\nonumber \\
{\cal L}^{\mu \nu}(-)&=& 2 i \mu \epsilon^{\mu\nu\delta\rho} q_{\delta}
(s_{l})_{\rho}.
\end{eqnarray}

In the QCD-based models, when the gluons from the Pomeron couple
to a single quark in the hadron, the effective Pomeron coupling
$V_{h I\hspace{-1.1mm}P}^{\mu} =\beta_{h I\hspace{-1.1mm}P}
\gamma^{\mu}$ appears which  looks like a $C= +1$ isoscalar
photon vertex \cite{lansh-m}.  The spin-dependent Pomeron
coupling can be obtained if one considers in the electromagnetic
nucleon current together with the Dirac form factor the Pauli
one \cite{nach}. We use in calculations the following form of
the two--gluon coupling with the proton \cite{golj}
\begin{equation}
V_{pI\hspace{-1.1mm}P}^{\mu\nu}(p,t,x_P)=
V_{pgg}^{\mu\nu}(p,t,x_P)=4 p^{\mu} p^{\nu}
A(t,x_P) +(\gamma^{\mu} p^{\nu} +\gamma^{\nu} p^{\mu}) B(t,x_P).
\label{ver}
\end{equation}
Here $x_P$ is a fraction of initial proton momenta carried by
the Pomeron. The term proportional to $B$ represents the
Pomeron coupling that leads to the non-flip amplitude. The $A$
function  is the spin--dependent part of the Pomeron coupling
that produces  spin--flip effects nonvanishing at high-energies
\cite{golj}. The absolute value of the ratio of $A$ to $B$ is
proportional to the ratio of helicity-flip and non-flip
amplitudes. It has been found in \cite{gol_mod,gol_kr} that
$\alpha=|A|/|B| \sim 0.1 -0.2 \,\mbox{GeV}^{-1}$ and has weak
energy dependence. We shall use in our estimations the value
$\alpha \leq 0.1 \,\mbox{GeV}^{-1}$.

 The hadronic tensor for the vertex (\ref{ver}) has the form
\begin{equation}
\label{wtenz}
W^{\alpha\alpha';\beta\beta'}(s_p)= \sum_{s_f} \bar u(p',s_f)
 V_{pgg}^{\alpha\alpha'}(p,t,x_P) u(p,s_p) \bar u(p,s_p)
V_{pgg}^{\star\,\beta\beta'}(p,t,x_P) u(p',s_f).
\end{equation}
Here  $s_p$ is a spin of the initial proton. The  spin-average
and the spin dependent hadronic tensor $W(\pm)$ is determined by
\begin{equation}
W^{\alpha\alpha';\beta\beta'}(\pm)=\frac{1}{2}( W^{\alpha\alpha';\beta\beta'}
(+\frac{1}{2}) \pm W^{\alpha\alpha';\beta\beta'}(-\frac{1}{2})),
\end{equation}
where $W(\pm\frac{1}{2})$ are the hadron tensors with the
helicity of the initial proton equal to $\pm 1/2$. The leading term
of the spin average hadron tensor looks like
\begin{equation}
W^{\alpha\alpha';\beta\beta'}(+) = 16 p^{\alpha} p^{\alpha'} p^{\beta}
 p^{\beta'} ( |B(t)+2 m A(t)|^2 + |t| |A(t)|^2).
\end{equation}
It is proportional to the meson-proton cross section up to a
function of $t$ ($m$ is a proton mass). The  spin-dependent part
of the hadron tensor can be written as
\begin{equation}
\label{w-f1}
W^{\alpha\alpha';\beta\beta'}(-) = \Delta
A_{\gamma}^{\alpha\alpha';\beta\beta'} +
\Delta A_{1}^{\alpha\alpha';\beta\beta'}.
\end{equation}
Here
\begin{eqnarray}
\label{af}
&&\Delta A_{\gamma}^{\alpha\alpha';\beta\beta'}=2 i m |B|^2\cdot \nonumber\\
&&\left[
p^{\alpha'} p^{\beta'} \epsilon^{\alpha\beta\gamma\delta}
(r_P)_\gamma (s_p)_\delta
 +\left(^{\mbox{All}\ \
\mbox{Per-}} _{\mbox{mutations}}\right)
\left(^{\{\alpha \to \alpha' \}} _{\{\beta \to \beta' \}}\right)
 \right]
\end{eqnarray}
The $\Delta A_{\gamma}$ contribution is proportional to $|B|^2$.
It is equivalent in form to the spin-dependent part of the leptonic tensor (see
(\ref{lpm})) and caused by the $\gamma_{\mu}$- term in (\ref{ver}). The
$\Delta A_{1}$ term contains interference of the $A$ and $B$ amplitudes from
(\ref{ver}). It is more complicated, and its explicit form can be
found in \cite{golj}.

\section{Spin effects in heavy $Q \bar Q$ production at COMPASS}
The diffractive  $Q \bar Q$ production in the lepton-proton
reaction is determined in the D-L model by the diagram of Fig. 1a.
Otherwise, in the two-gluon picture of the Pomeron we consider
all the graphs where the gluons from the Pomeron couple to a
different quark in the loop (Fig. 1c) as well as to the single one
(Fig. 1b). This provides a gauge-invariant scattering
amplitude. The spin-average and spin-dependent cross section can
be written in the form
\begin{equation}
\label{sigma}
\frac{d^5 \sigma(\pm)}{dQ^2 dy dx_p dt
dk_\perp^2}=
\left(^{(2-2 y+y^2)} _{\hspace{3mm}(2-y)}\right)
 \frac{C(x_p,Q^2) \; N(\pm)}
{\sqrt{1-4k_\perp^2\beta/Q^2}}.
\end{equation}
Here $C(x_p,Q^2)$ is a normalization function which is common
for the spin average and spin dependent cross section; $N(\pm)$
is determined by the sum of graphs in Fig. 1 b,c integrated over
the gluon momenta $l$ and $l'$.
\begin{eqnarray}
\label{tpm}
N(\pm) = \int \frac{ d^2 l_{\bot} d^2 l_{\bot}'
D^{\pm}(t,Q^2,l_{\bot},l'_{\bot},\cdots)}
{(l_\perp^2+\lambda^2)((\vec l_\perp+\vec
r_{\perp})^2+\lambda^2) (l_\perp^{'2}+\lambda^2)((\vec l'_\perp+\vec
r_{\perp})^2+\lambda^2)}
\end{eqnarray}
The $D^{\pm}$ function here is a sum of traces over the quark
loops of the graphs in Fig. 1b,c and corresponding crossed
diagrams convoluted with the spin average and spin-dependent
tensors. The calculation shows a considerable cancellation
between the nonplanar contribution of the graph in Fig. 1c and
the planar contribution from Fig. 1b. As a result, the function
$D^{\pm}$ in (\ref{tpm}) is proportional to the gluon momenta
$l_{\perp}$ and $l'_{\perp}$. This distinguishes the D-L and the
QCD models of the Pomeron.  The D-L Pomeron contribution in Fig.
1a is equivalent only to the planar graph of Fig 1b.

We shall discuss here the $A_{ll}=\sigma(-)/\sigma(+)$ asymmetry
at COMPASS energy.  The obtained asymmetry is proportional to
$x_p$ ($x_p$ is typically of about $.05-.1$)  and has a weak
energy dependence. The predicted asymmetry is quite small and
does not exceed 1-1.5\%. It has a week dependence on the
$\alpha=A/B$ ratio and does not vanish for $\alpha=0$. The $Q^2$
dependence of the $A_{ll}$ asymmetry of the diffractive open
charm production can be estimated as $A_{ll} \propto
Q^2/(Q^2+Q^2_0)$ and is shown in Fig. 2. The COMPASS experiment
intends to study events at small $Q^2$ where the diffractive
asymmetry will be extremely small. Thus, our calculations within
two-gluon model of the Pomeron shows that there is no problem
with the diffractive contribution at COMPASS. Note that the
smallness of the asymmetry is caused mainly by the strong
cancellation in $\Delta \sigma$ between the graphs in Fig 1b,c.
Such compensation is absent in the D-L model where only the
planar graphs shown in Fig. 1a (Fig. 1b) appear. As a result,
the latest model of the Pomeron, provides the diffractive
$A_{ll}$ asymmetry which is about 10\% \cite{gola_ll} and is
larger by a factor of about 10 than the value obtained here.
\\[.4cm]
\begin{minipage}{6cm}
\epsfxsize=6.0cm
\centerline{\epsfbox{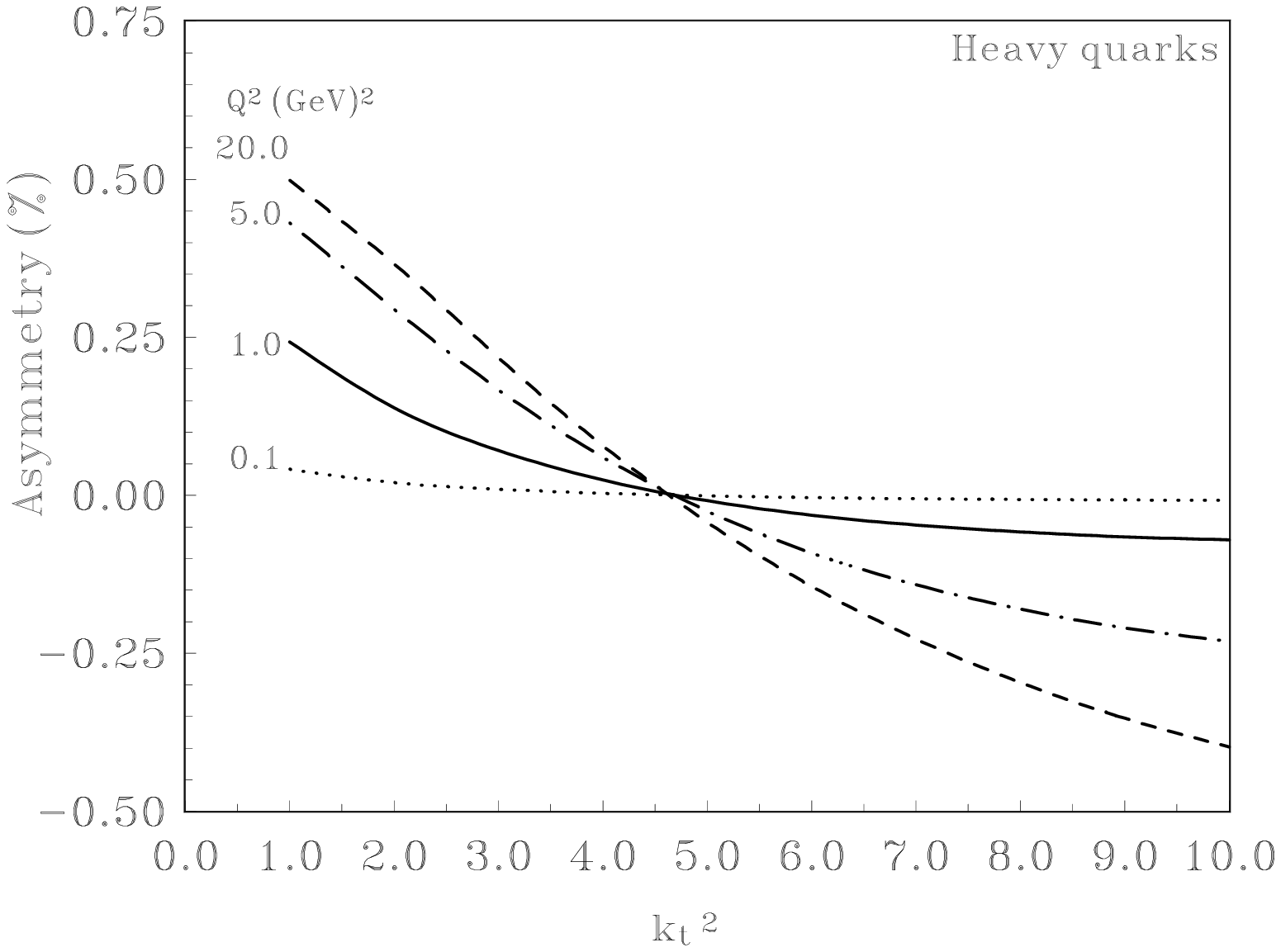}}
\end{minipage}
\begin{minipage}{0.14cm}
\phantom{aa}
\end{minipage}
\begin{minipage}{6cm}
\epsfxsize=6.0cm
\centerline{\epsfbox{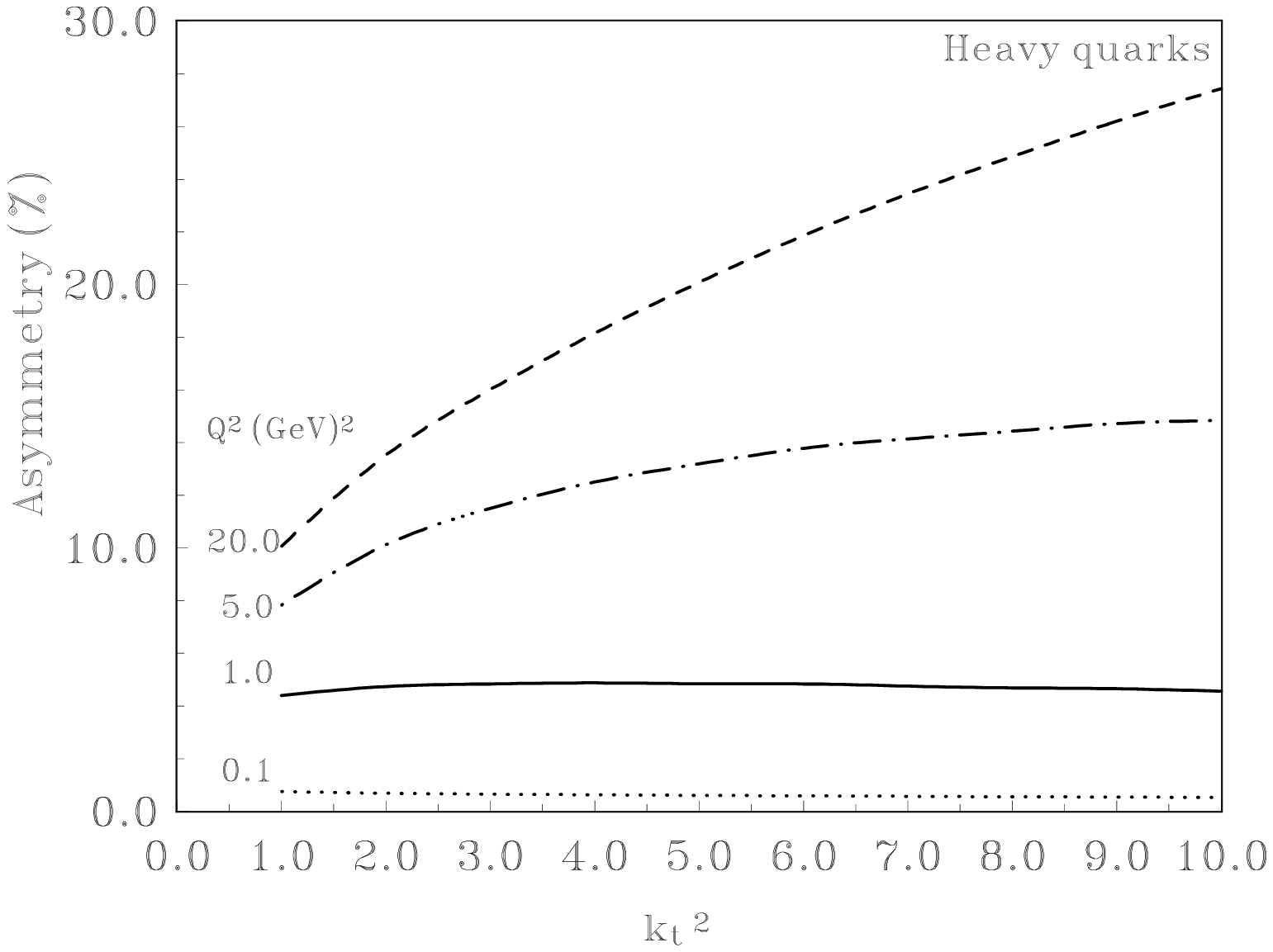}}
\end{minipage}
\\[.3cm]
\noindent
\begin{minipage}{6cm}
Fig. 2.~ {The predicted $Q^2$ dependence of the $A_{ll}$ asymmetry
for the $c \bar c$ production at COMPASS for
$\alpha=0.1 \mbox{GeV}^{-1}$, $x_p$=0.1, $y$=0.5}.
\end{minipage}
\begin{minipage}{.15cm}
\phantom{aa}
\end{minipage}
\begin{minipage}{6cm}
Fig. 3.~ {The predicted $Q^2$ dependence of the $A_{lT}$ asymmetry
for the $c \bar c$ production at COMPASS for
$\alpha=0.1\mbox{GeV}^{-1}$, $x_p$=0.1, $y$=0.5}.
\end{minipage}
\\[.4cm]
The other important object which can be studied at COMPASS is
the $A_{lT}$ asymmetry with longitudinal lepton and transverse
proton polarization. It has been found that the $A_{lT}$ asymmetry
is proportional to the scalar production of the proton spin
vector and the jet momentum $A_{lT} \propto (s_{\perp} \cdot
k_{\perp}) \propto \cos (\phi_{Jet})$. Thus, the asymmetry
integrated over the azimuthal jet angle $\phi_{Jet}$ is zero. We
have calculated the $A_{lT}$ asymmetry for the case when the
proton spin vector is perpendicular to the lepton scattering
plane and the jet momentum is parallel to this spin vector. The
predicted asymmetry is show in Fig. 3. It is huge and has a
drastic $k^2_{\perp}$ dependence. The large value of the $A_{lT}$
asymmetry is caused by the fact that in contrast to $A_{ll}$, it
does not have a small factor $x_p$ as a coefficient.

Note that usually it is impossible to detect the outgoing proton
in fixed target experiments. Then, the integration
over the momentum transfer $t$ of the cross section
(\ref{sigma}) should be done. Such integrated asymmetry is
smaller by the factor of 1.2--2 with respect to the unintegrated
values shown in Figs 2, 3.

\section{Spin effects in vector meson production}

Similar calculations have been done for the diffractive  vector
meson leptoproduction. The $\gamma^\star  I\hspace{-1.6mm}P \to
V$ transition has been modeled as a conversion of the virtual
photon in a $q \bar q$ pair and subsequent $q \bar q \to V$
transition. As previously, we have included in our analysis the
graphs where the gluons from the Pomeron couple to a different
quark in the loop as well to the single one. The nonrelativistic
wave function, when both the quarks have the same momenta equal
to half of the meson momentum and the quark mass $m$ is equal to
$m_V/2$, have been used in calculation. The spin-dependent cross
section  can be written in the form
\begin{equation}
\frac{d\sigma^{\pm}}{dQ^2 dy dt}=\frac{|T^{\pm}|^2}{32 (2\pi)^3
 Q^2 s^2 y}. \label{ds}
\end{equation}
For the spin-average  amplitude square we find \cite{golj}
\begin{equation}
 |T^{+}|^2=  N ((2-2 y+y^2) m_J^2 + 2(1 -y) Q^2) s^2 |B|^2
[|1+2 m \alpha|^2+|\alpha|^2 |t|] I^2.
\label{t+}
\end{equation}
Here $N$ is a known normalization factor $\alpha=|A|/|B|$ and
$I$ is the integral over transverse momentum of the gluon. The
term proportional to $(2-2 y+y^2) m_J^2$ in (\ref{t+})
represents the contribution of a virtual photon  with transverse
polarization. The $2(1 -y) Q^2$ term describes the effect of
longitudinal photons.
\\[.5cm]
\begin{minipage}{6cm}
\epsfxsize=5.8cm
\centerline{\epsfbox{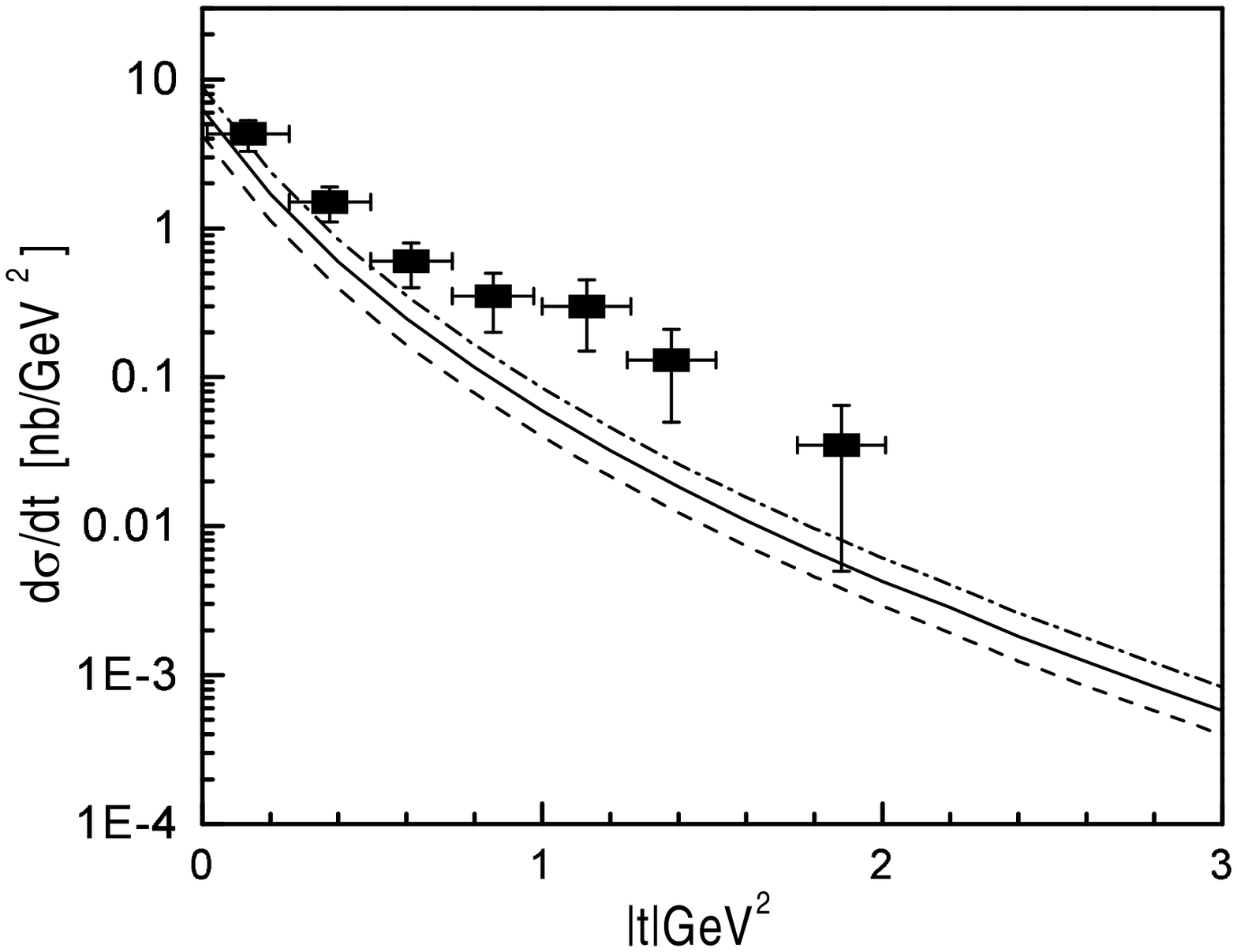}}
\end{minipage}
\begin{minipage}{0.14cm}
\phantom{aa}
\end{minipage}
\begin{minipage}{6cm}
\epsfxsize=5.8cm
\centerline{\epsfbox{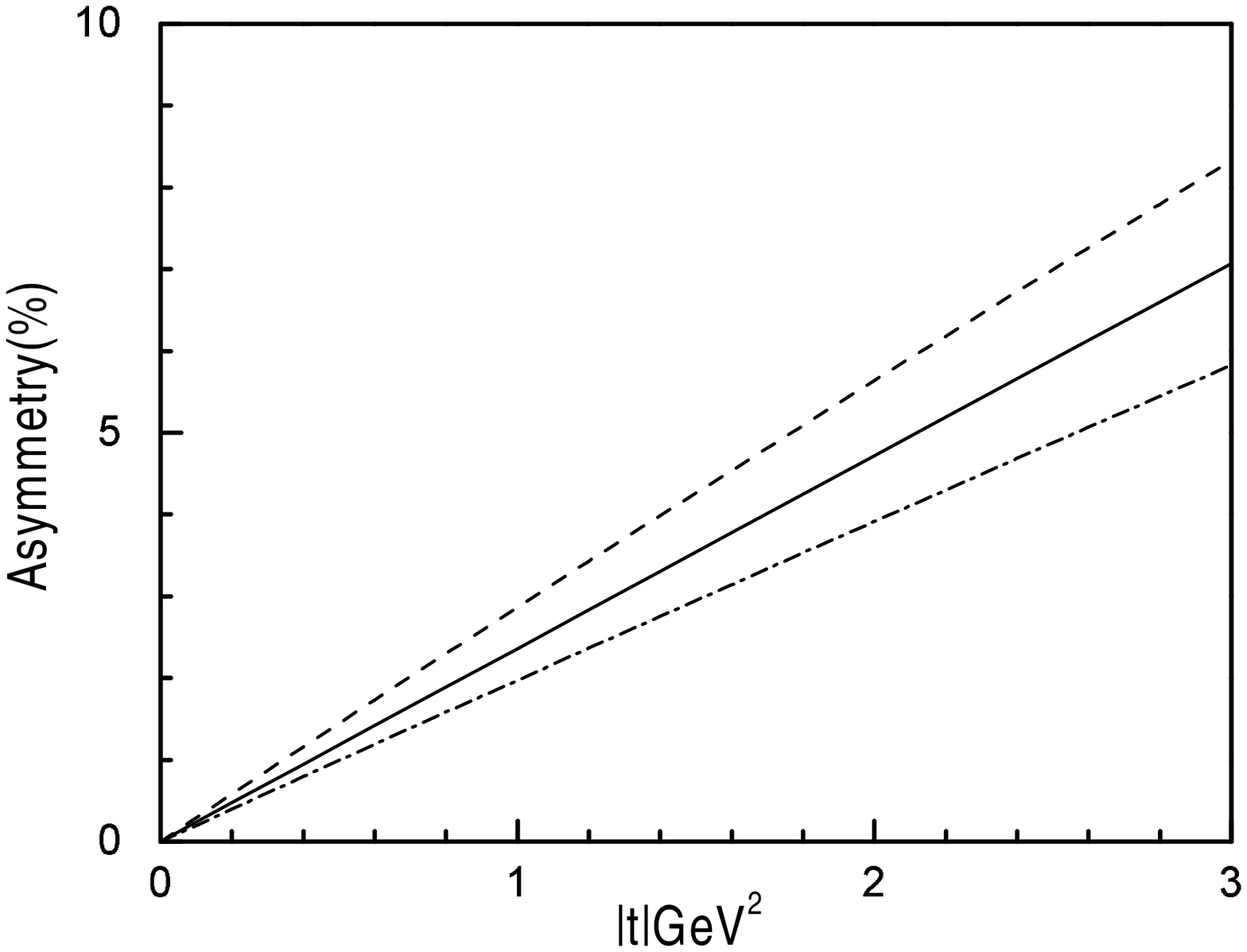}}
\end{minipage}
\\[3mm]
\begin{minipage}{6cm}
Fig. 4.~ {The differential cross section of the $J/\Psi$ production
at HERA energy: solid line -for $\alpha=0$; dot-dashed line -for
$\alpha=0.1\mbox{GeV}^{-1}$;
dashed line -for $\alpha=-0.1\mbox{GeV}^{-1}$}.
\end{minipage}
\begin{minipage}{.15cm}
\phantom{aaa}
\end{minipage}
\begin{minipage}{6cm}
Fig. 5.~ {The predicted $A_{ll}$ asymmetry of the $J/\Psi$ production
at COMPASS: solid line -for $\alpha=0$; dot-dashed line -for
$\alpha=0.1\mbox{GeV}^{-1}$;
dashed line -for $\alpha=-0.1\mbox{GeV}^{-1}$}.
\end{minipage}
\\[4mm]
The cross sections (\ref{ds}) integrated over $y$ and $Q^2$ with
$Q^2_{max} \sim 4 \mbox{GeV}^2$
\begin{equation}
\frac{d\sigma^{\pm}}{dt}=\int_{y_{min}}^{y_{max}} dy
\int_{Q^2_{min}}^{Q^2_{max}} dQ^2
\frac{d\sigma^{\pm}}{dQ^2 dy dt}.
\end{equation}
for the $J/\Psi$ production at HERA energy
$\sqrt{s}=300GeV^2$
is shown in Fig. 4.  The spin-average cross sections
are sensitive to $\alpha$ but the shape of all curves is the
same. Thus, it is difficult to extract information about the
spin--dependent part of the Pomeron coupling from the
spin--average cross section of the diffractive vector meson
production.

The spin-dependent amplitude square looks like
\begin{equation}
 |T^{-}|^2= N (2- y)  s |t|  [|B|^2+ 2 m |A B| ] m_J^2 I^2.
\label{t-}
\end{equation}
 As a result, the following form of
asymmetry is found \cite{golj}:
\begin{equation}
\label{all_a}
A_{ll}= \frac{\sigma(-)}{\sigma(+)} \sim \frac{|t|}{s}\frac{(2-\bar y)
(1+2 m \alpha)}
{(2-2\bar y+\bar y^2)\left[ (1+2 m \alpha)^2+\alpha^2|t| \right]
}.
\end{equation}
The predicted asymmetry of the $J/\Psi$ vector meson production
at  COMPASS for different $\alpha$ is shown in Fig.\ 5. The
asymmetry is equal to zero in the forward direction. The
predicted asymmetry  does not vanish for nonzero $|t|$.  The
value of the asymmetry for $\alpha \neq 0$ is dependent on  the
$A$--term of the Pomeron coupling. However, the sensitivity of
the asymmetry to $\alpha$ is quite weak. Thus, it will not be so
easy to study the spin structure of the Pomeron coupling with
the proton from the $A_{ll}$ asymmetry of the diffractive vector
meson production. The obtained asymmetry is independent of the
mass of a produced meson. We can expect a similar value of the
asymmetry in the polarized diffractive $\phi$ --meson
leptoproduction. In this reaction the contribution of the $t$
-channel $Q \bar Q$ exchange should be quite small. Otherwise,
for the $\rho$ meson production the $t$ -channel $Q \bar Q$
contribution should be significant at COMPASS energies.

\section{Conclusion}
In the present report, the polarized cross section of the
diffractive hadron leptoproduction at high energies has been
studied.  As a result, connection of the spin--dependent cross
section in the diffractive production with the Pomeron coupling
has been found. Generally, two--gluon couplings with the proton
in diffractive processes at small $x$ can be expressed in terms
of the skewed gluon distribution in the nucleon ${\cal
F}_X(X+\Delta)$, where for vector meson production one can find
$X\sim (Q^2+m_V^2)/W^2, \Delta \ll X$ \cite{rad}. Here
$X+\Delta$ is a fraction of the proton momentum carried by the
outgoing gluon, and the difference between the gluon momenta
(skewedness) is equal to $X$. The function $B$ should be
determined by the spin--average and the function $A$ by the
polarized skewed gluon distribution in the proton.  To find the
explicit connection of $A$ with spin--dependent gluon
distribution, additional study is necessary.

It has been found that there in no problem with the diffractive
contribution to $A_{ll}$ at COMPASS for $Q^2 \to 0$ because its
value is found to be negligible. The predicted $A_{ll}$
asymmetry in the $Q \bar Q$ leptoproduction for $Q^2 \geq 1
\mbox{GeV}^2$ is smaller than 1.5\%. The nonzero asymmetry for
$\alpha=A/B=0$  is completely determined by the
$\gamma^{\alpha}$ term in the Pomeron coupling (\ref{ver}). The
$A_{ll}$ asymmetry in diffractive processes for nonzero momentum
transfer has been found dependent on the $A$ term of the
Pomeron coupling which has a spin dependent nature. However, the
sensitivity of asymmetry to the $\alpha$--ratio  is quite weak.
Thus, the $A_{ll}$ asymmetry in diffractive reactions is not a
good tool to study the polarized gluon distributions of the
proton and spin structure of the Pomeron. Otherwise, the
$A_{lT}$ asymmetry in diffractive $Q \bar Q$ production might be
about 10-20\%. This asymmetry is proportional to $\alpha$ and
might be used to obtain direct information about the spin
structure of Pomeron coupling and the skewed polarized gluon
structure functions of the proton.

We would like to stress that the spin-dependent amplitude in the
diffractive reaction is more sensitive to the model of the
Pomeron interaction than the spin-average one. Some tricks in
calculations, which restore the gauge invariance of the
scattering amplitude, can simplify investigation of $\sigma(+)$ (see
e.g \cite{habeck}). Unfortunately, they do not work for
$\sigma(-)$. We consider the full perturbative
calculation which includes all possible $t$-channel gluon
exchanges as the most reliable method to study
the spin effects in diffractive reactions.
Future polarized diffractive experiments might be an important
test of the spin structure of QCD at large distances and of the
different theoretical approaches to diffractive reactions.

The author is grateful to the Organizing Committee for financial support.

\end {document}